\newcommand{\apjl}{ApJL}

\newcommand{\aap}{A\&A}

\documentclass[11pt]{article}

\usepackage{wds11,epsf}
\usepackage{natbib}
\usepackage{url}
\usepackage{graphicx}
\usepackage{amsmath}
\usepackage{amsthm}
\usepackage{wasysym}
\usepackage{epstopdf}
\usepackage{natbib}
\usepackage{multirow}
\usepackage{booktabs}
\usepackage{fancyref}
\usepackage[unicode]{hyperref} 

\lefthead{ZAJA\v{C}EK ET AL.}
\righthead{Dusty S-cluster Object near Sgr~A*}

\begin{document}

\title{Infrared-excess Source DSO/G2 Near the Galactic Center: Theory vs. Observations}

\author{M. Zaja\v cek and A. Eckart}

\affil{University of Cologne, First Physics Institute, Cologne, Germany}
\affil{Max-Planck Institute for Radioastronomy, Bonn, Germany}

\author{F. Peissker and G. D. Karssen}

\affil{University of Cologne, First Physics Institute, Cologne, Germany}

\author{V. Karas}

\affil{Astronomical Institute of the Academy of Sciences of the Czech Republic}

\begin{abstract}
Based on the monitoring of the Dusty S-cluster Object (DSO/G2) during its closest approach to
the Galactic Center supermassive black hole in 2014 and 2015 with ESO VLT/SINFONI, we further explore
the model of a young, accreting star to explain observed spectral and morphological features.
The stellar scenario is supported by our findings, i.e. ionized-hydrogen emission from the DSO that
remains spatially compact before and after the peribothron passage.
The detection of DSO/G2 object as a compact single-peak emission-line source is not consistent
with the original hypothesis of a core-less cloud that
is necessarily tidally stretched, hence producing a double-peak emission line profile around the pericentre passage.
This strengthens the evidence that the DSO/G2 source is a dust-enshrouded young star that appears to be in an accretion phase. The infall of material from the circumstellar disc
onto the stellar surface can contribute significantly to the emission of Br$\gamma$ line as well as the observed large line width of the order of 10 angstrom. 
\end{abstract}

\begin{article}

\section{Introduction}

The centre of the Milky Way is the closest galactic nucleus whose physics, namely dynamical processes, can be studied with the high resolution facilities in near-infrared (NIR) bands using 8--10 m class telescopes equipped with adaptive optics (AO) systems \citep{2005bhcm.book.....E}. These instruments allow us to study the motion of high-velocity stars within the innermost arcsecond, the so-called S-stars of spectral type B \citep[][and references therein]{gillessen09}. All members of the S-cluster are tightly bound to the centrally concentrated dark mass of $4\times 10^6\,M_{\odot}$, whose position is in agreement with the compact radio source Sgr~A* associated with the supermassive black hole (SMBH) \citep[see][for a review]{2013CQGra..30x4003F}. S-stars appear to be ordinary B-type stars and probably formed recently, 6--400 Myr ago \citep{2005ApJ...628..246E}. The discovery and monitoring of NIR-excess objects in the S-cluster raised the possibility of an existence of even younger, pre-main-sequence stellar objects ($\lesssim 1\,\rm{Myr}$) that are still in an accretion phase \citep{eck13,val15}.

The fast-moving infrared excess source named Dusty S-cluster Object (DSO; also called G2) within the cluster of high-velocity S-stars was discovered by \citet{gil12} and
subsequently monitored and analyzed \citep{eck13,gil13a,gil13b,wit14,pfuhl15,val15} to detect a potential effect on the activity of
the supermassive black hole (hereafter SMBH) as well as to determine the character of the object. Independent analysis of the available data by different groups has led to different viable interpretations of DSO/G2 nature \citep[see][for a review]{2015Natur.524..301B}.    

The main debate concerning the nature of the object is whether DSO/G2 is a core-less gaseous-and-dusty cloud or whether its compact form is instead supported by the presence of an enshrouded star inside an extended envelope.
While a core-less cloud of a few Earth masses \citep[][]{gil12} would be necessarily tidally stretched and would be expected to interact strongly with the surrounding medium,
a star should instead keep its compact nature and continue to orbit along the original trajectory. The orbital elements of the DSO are currently constrained very well;
see \citet{val15} for a recent fit. The source has passed the pericentre in 2014 at the distance of $r_{{\rm p}}=a(1-e)=0.033\,{\rm pc}\times (1-0.976) \approx 2000\,r_{\rm{s}}$,\
where $r_{\rm{s}}$ is the Schwarzschild radius ($r_{\rm{s}}\equiv 2GM_{\bullet}/c^2 \approx 2.95\times 10^5 M_{\bullet}/M_{\odot}\,\rm{cm}$;
we consider $M_{\bullet}=4\times 10^6\,M_{\odot}$ for the SMBH mass associated with the compact radio source Sgr~A*).
Any stellar object associated with the DSO is not tidally disrupted, since the critical tidal radius for
the SMBH is $r_{\rm{t}}=R_{\star}(3M_{\bullet}/M_{\star})^{1/3} \approx 13 (R_{\star}/R_{\odot})(M_{\star}/M_{\odot})^{-1/3}\,r_{\rm{s}}$,
where $R_{\star}$ is the stellar radius. However, the tidal radius for an extended,
unbound cloud with the radius of $R_{\rm{cl}}\approx 30\,\rm{mas}\approx 5\times 10^4\,R_{\odot}$ and
the mass of three Earth masses ($\sim 10^{-5}\,M_{\odot}$) \citep{gil12} is $\sim 3\times 10^{7}\,r_{\rm{s}}$, hence it should undergo
the full tidal disruption at the pericentre. Thus, there is a clear distinction between the star and the cloud concerning the orbital stability.
The predictions of evolution for different scenarios and geometries were numerically analysed in \citet{zajacek14} and references therein. The basic qualitative results of the paper are in agreement with analytical estimates: the core-less
cloud loses angular momentum and its trajectory starts to progressively deviate from a closed elliptic orbit after the pericentre passage. The star, on the other hand, continues to orbit 
the SMBH along its original trajectory. 

In this contribution we further elaborate on the model of a dust-enshrouded, presumably young star \citep{murray-clay12,scoville13,ballone13,zajacek14,deColle14},
focusing mainly on the plausible contribution of the inflow to the overall emission of the DSO.
The structure of the paper is as follows.
In Section \textit{Observations of the DSO and other NIR-excess sources} we summarize the main observational properties of the DSO and
other similar sources located in the inner $2''$ from the Galactic centre. Consequently, in Section \textit{DSO as a young star}
we propose the model of a young star that can explain the broad hydrogen emission lines and the compact nature of the source observed so far.
Finally, we summarize the results in \textit{Conclusions}.

\section{Observations of the DSO and other NIR-excess sources}
\label{sec_obs}

The Dusty S-cluster object was discovered as a prominent L'-band source \citep{gil12}. Consequently, its Br$\gamma$ emission was detected and used for the update of the orbital elements \citep{gil13a,phi13}. Moreover, $K_{s}$-band detection was reported by \citet{eck13,2014IAUS..303..269E}, which agrees with the position of $L'$-band source as well as the source of Br$\gamma$ emission. $K_{s}$-band identification of the DSO is more consistent with its stellar nature rather than a core-less cloud. \citet{eck13} did spectral decomposition of the source using M-band measurement by \citet{gil12} that puts an upper limit of $\sim 30\,L_{\odot}$ on the luminosity of the source. The observations using $H$-, $K_{s}$-, and $L'$-band show that the source has an infrared excess of $H-K_{s}>2.3$ and $L'-K_{s}=4.5$ \citep{eck13}. The continuum measurements in $H-$, $K_{s}-$, and $L'-$ bands can be fitted either by the presence of exceptionally warm dust component of $550$--$650\,{\rm{K}}$ or by the presence of a stellar source embedded in the dust having the temperature of $\sim 450\,\rm{K}$ \citep[see][for a detailed discussion]{eck13,2014IAUS..303..269E}. 

\begin{figure}[tbh]
  \centering
  \includegraphics[width=\textwidth]{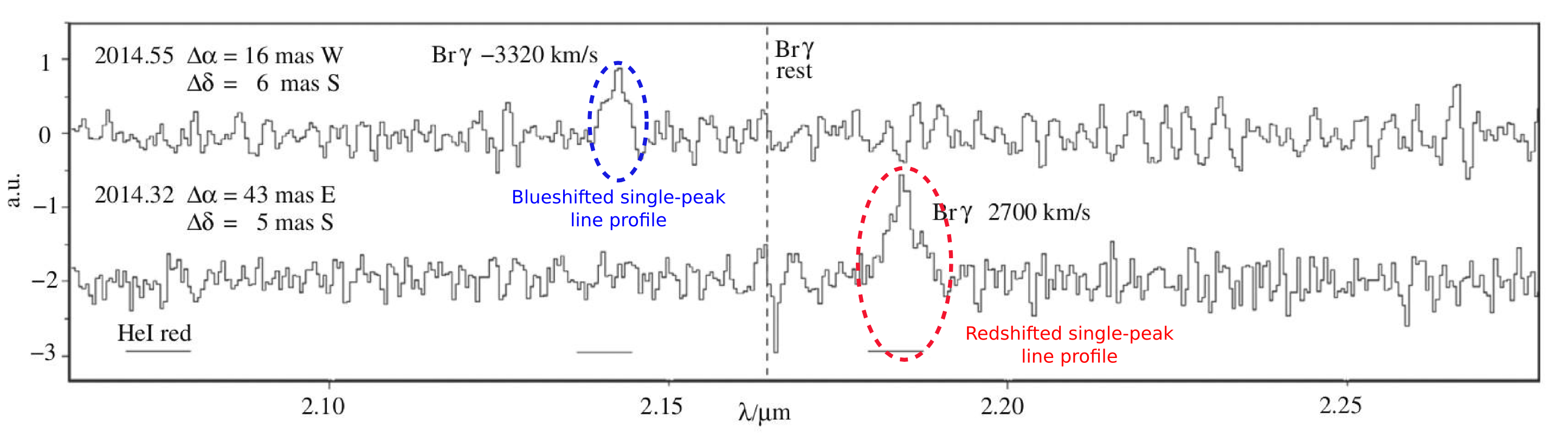}
  \caption{A NIR spectrum around Br$\gamma$ rest wavelength ($2.17\,\mu{\rm m}$). The bottom part shows the pre-peribothron single-peak, redshifted Br$\gamma$ emission, whereas the top part the post-pericentre single-peak, blue-shifted emission of Br$\gamma$. Figure adopted from \citet{val15}.}
  \label{fig_nir_spectrum}
\end{figure}

The observations of the DSO around the pericentre passage have been crucial for determining the compactness and thus the nature of the source. A near-infrared  analysis using integral field spectroscopy at $1.45$--$2.45\,{\rm \mu m}$ (ESO VLT/SINFONI) performed by \citet{val15} show that the DSO is a single-peak, emission line source both before and after the pericentre, see Fig. \ref{fig_nir_spectrum}. At the epoch of 2014.32 the source is $43\,{\rm mas}$ east of Sgr~A* and its line-of-sight velocity  corresponding to the line centroid is $+2700 \pm 60\,{\rm km\,s}^{-1}$. The blueshifted emission at the epoch of $2014.55$ is detected at $16\,{\rm mas}$ west of Sgr~A* and the radial velocity of the source is $-3320 \pm 60\,{\rm km\,s}^{-1}$. Hence, the source stays compact and the bulk of the emission comes within the region of $\sim 20\,{\rm mas}$ around the line peak.

\begin{figure}
  \begin{tabular}{cc}
      \includegraphics[width=0.5\textwidth]{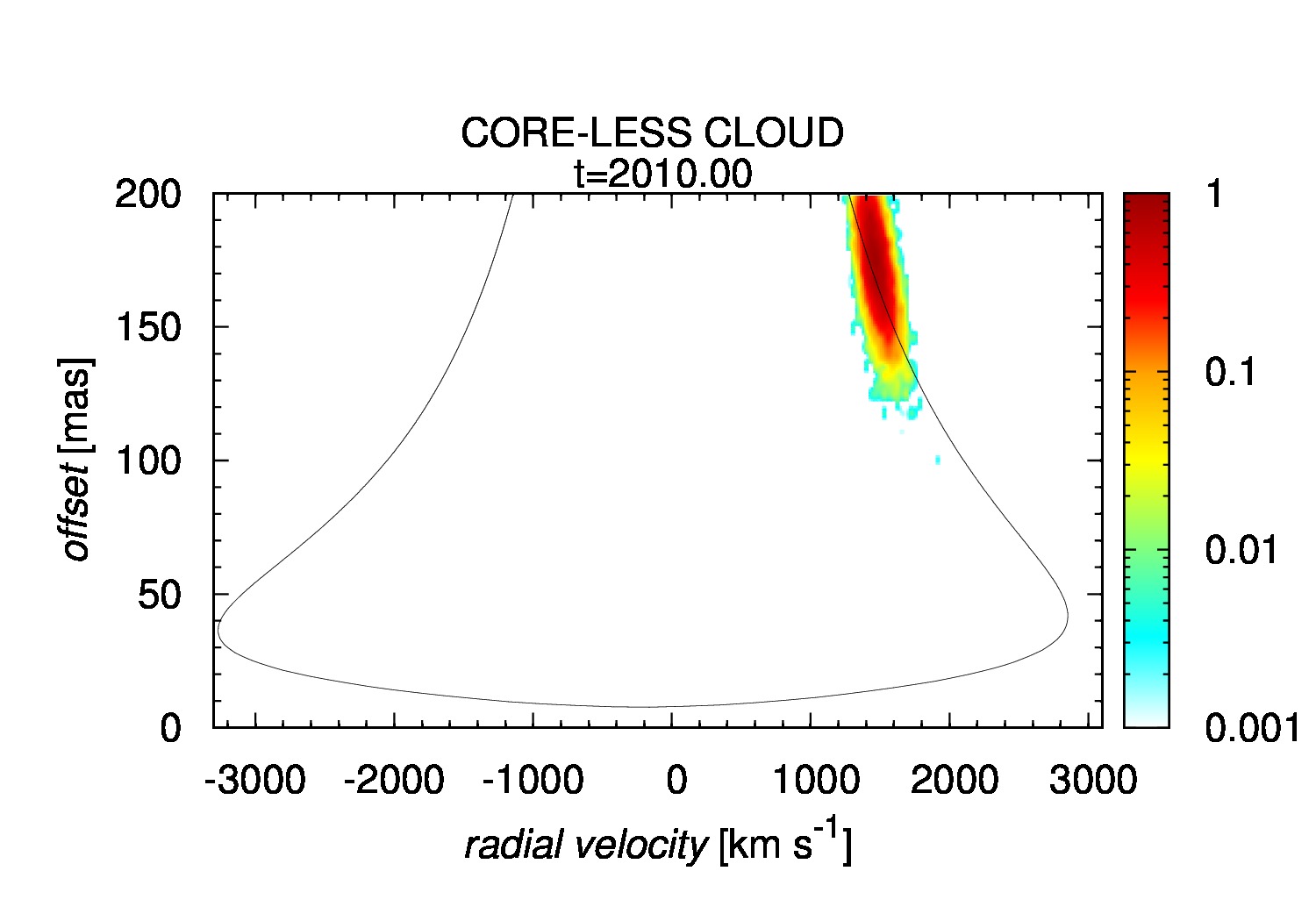} & \includegraphics[width=0.5\textwidth]{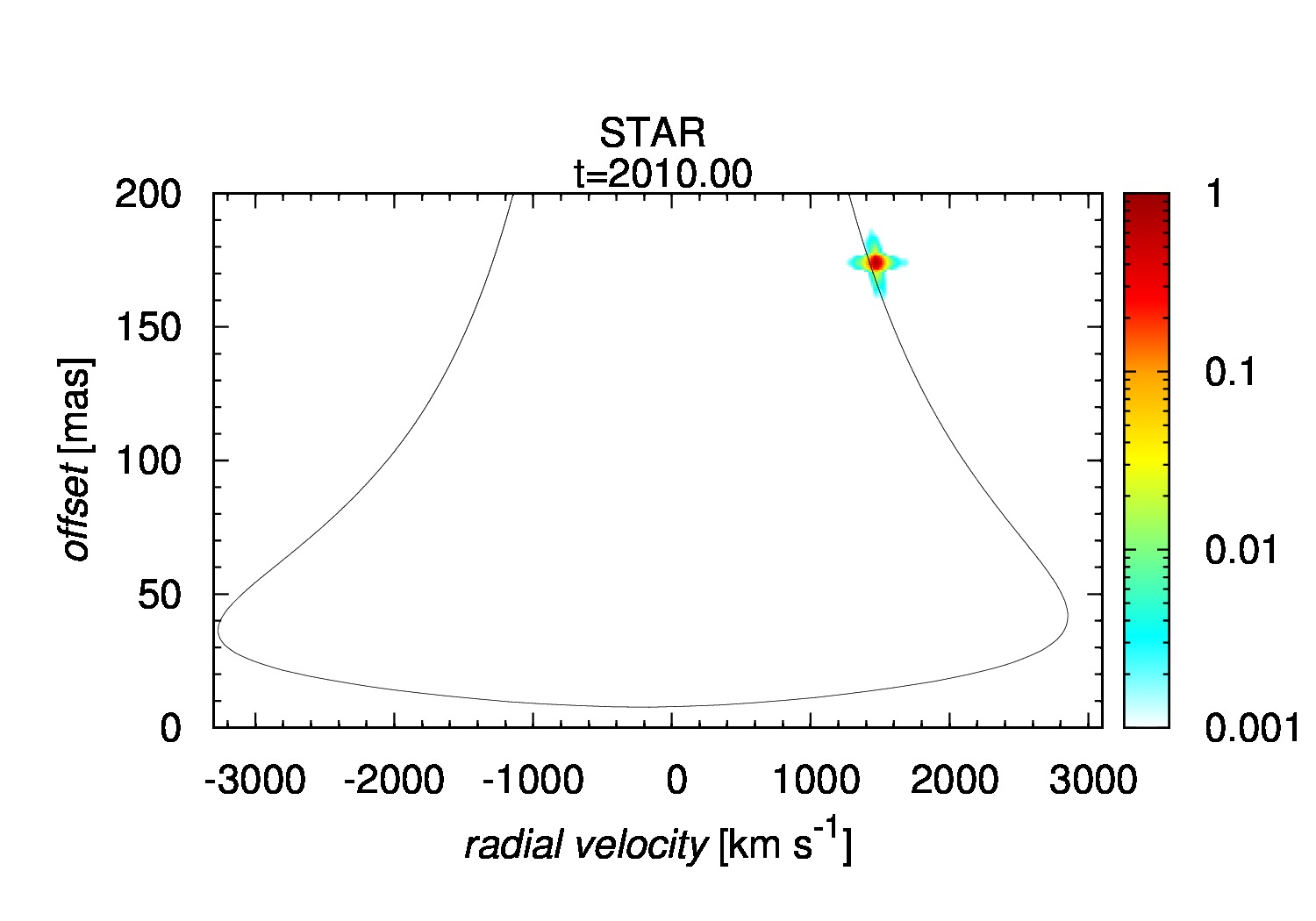}\\
       \includegraphics[width=0.5\textwidth]{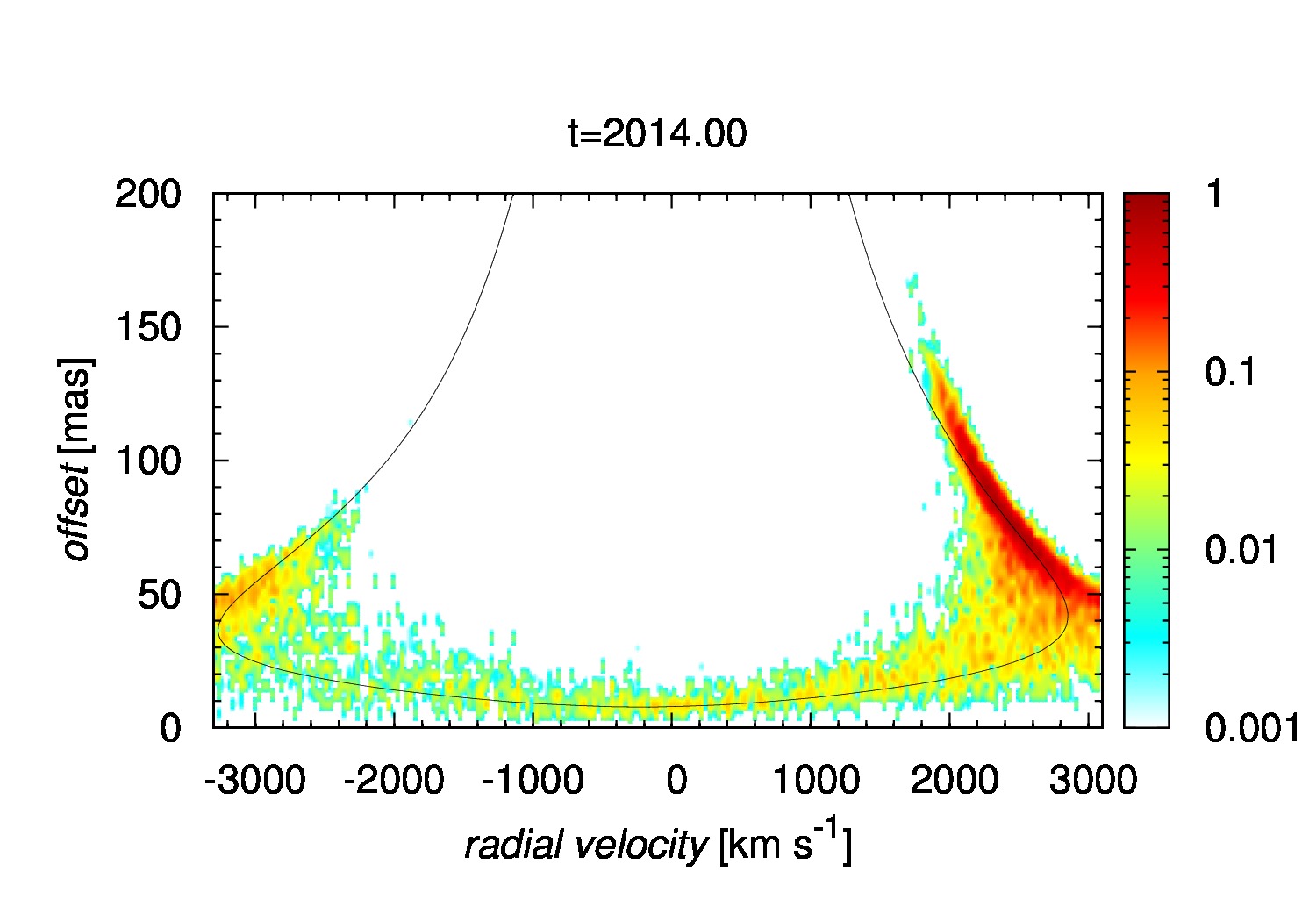} & \includegraphics[width=0.5\textwidth]{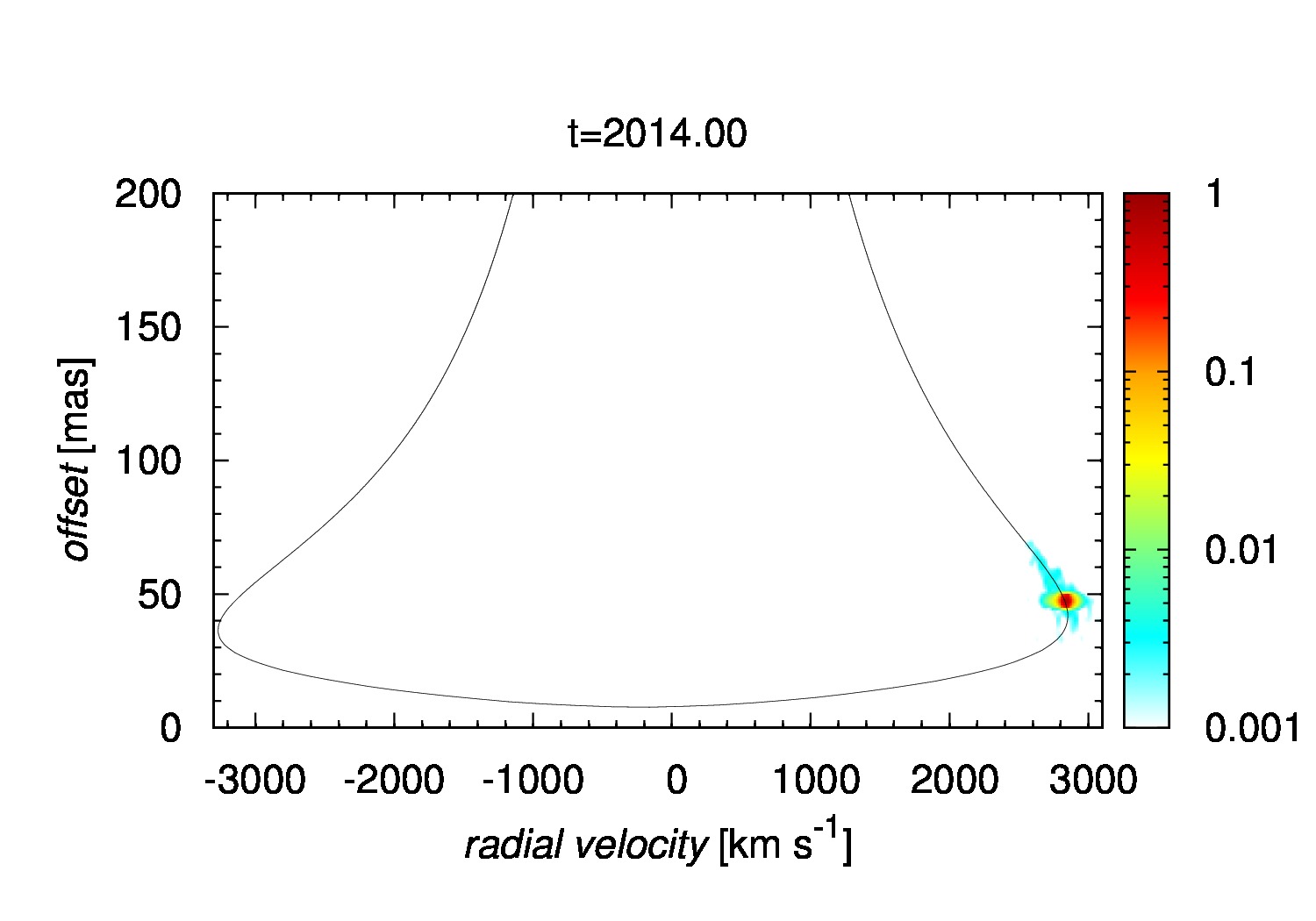}\\
        \includegraphics[width=0.5\textwidth]{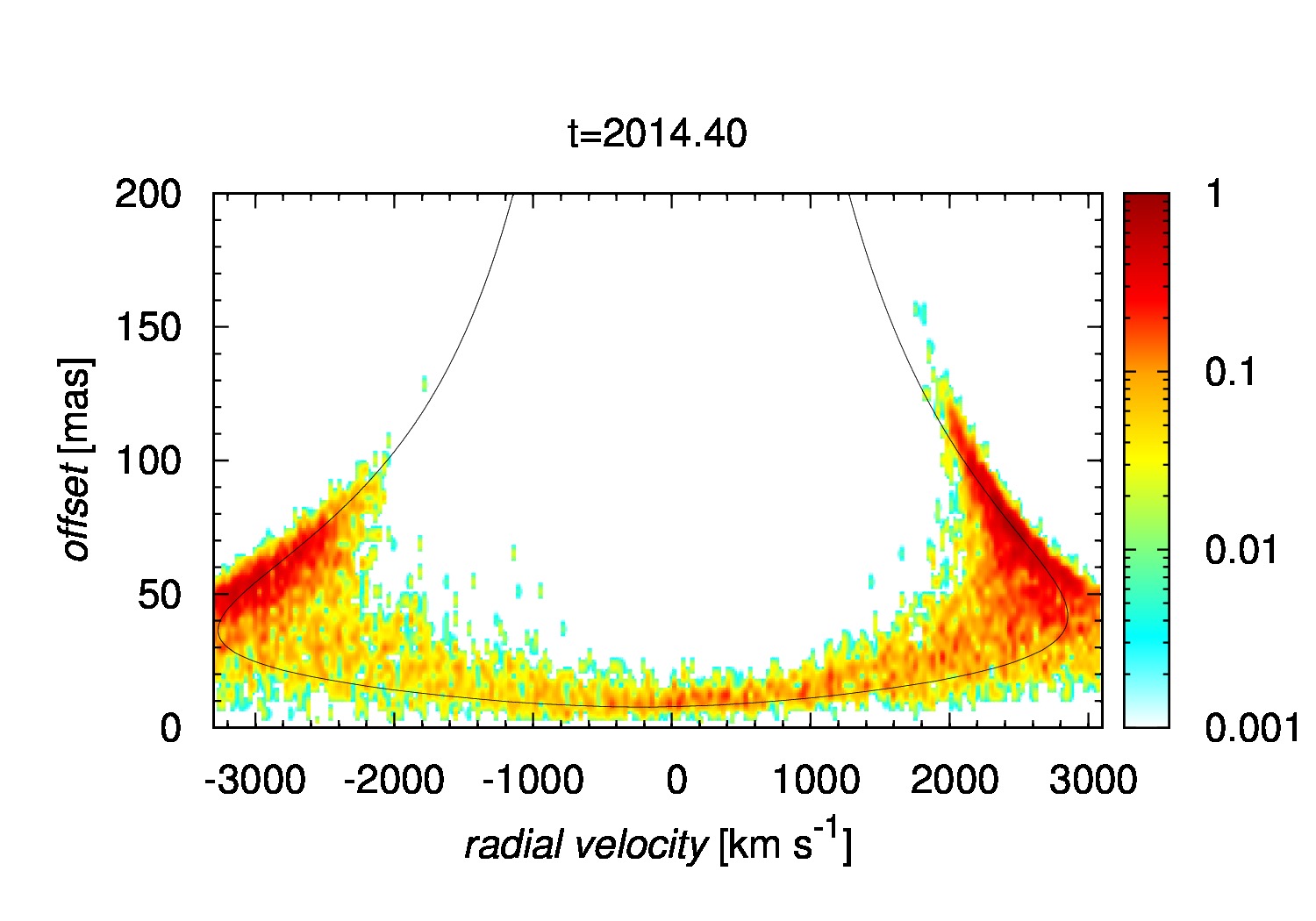} & \includegraphics[width=0.5\textwidth]{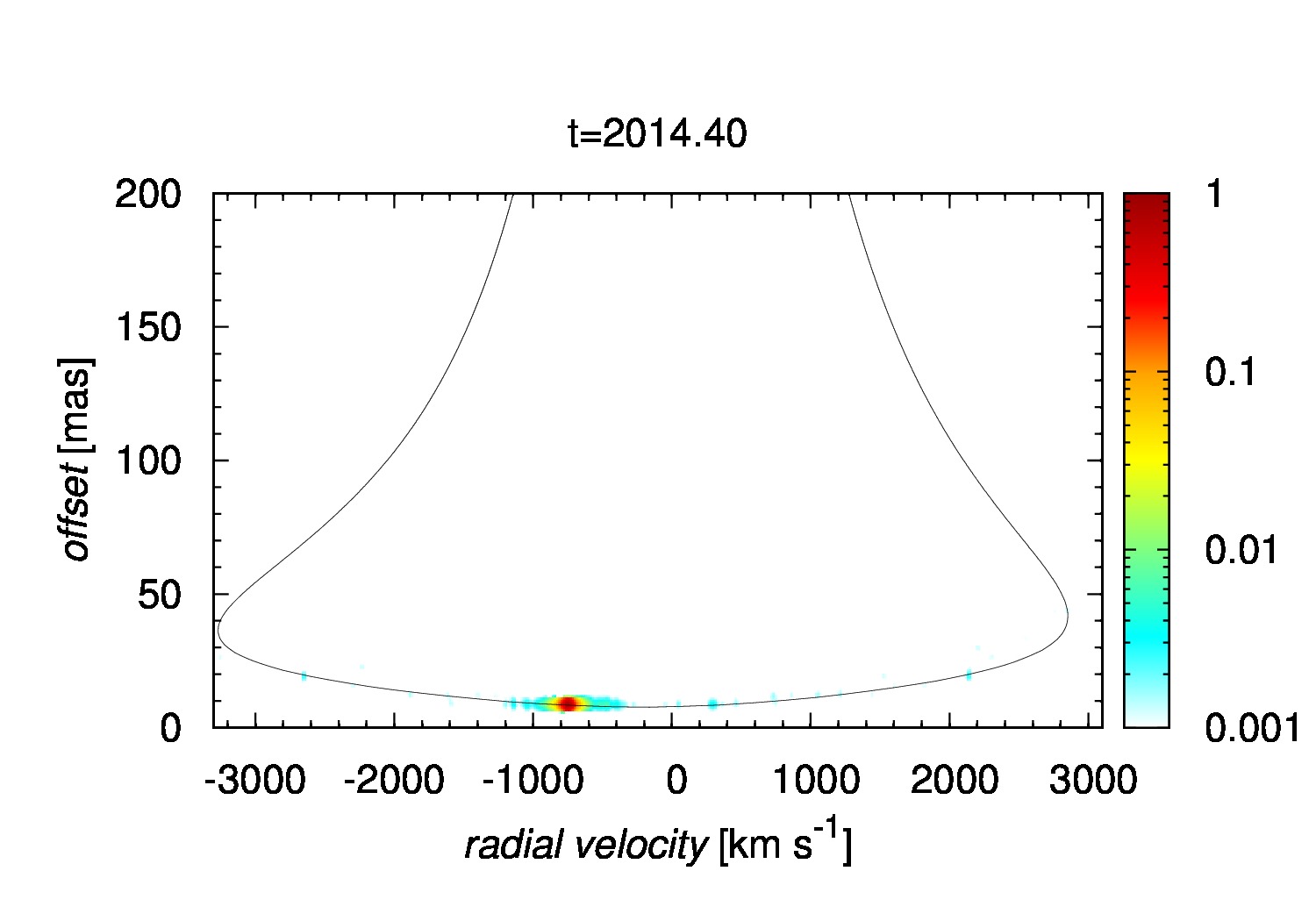}\\
         \includegraphics[width=0.5\textwidth]{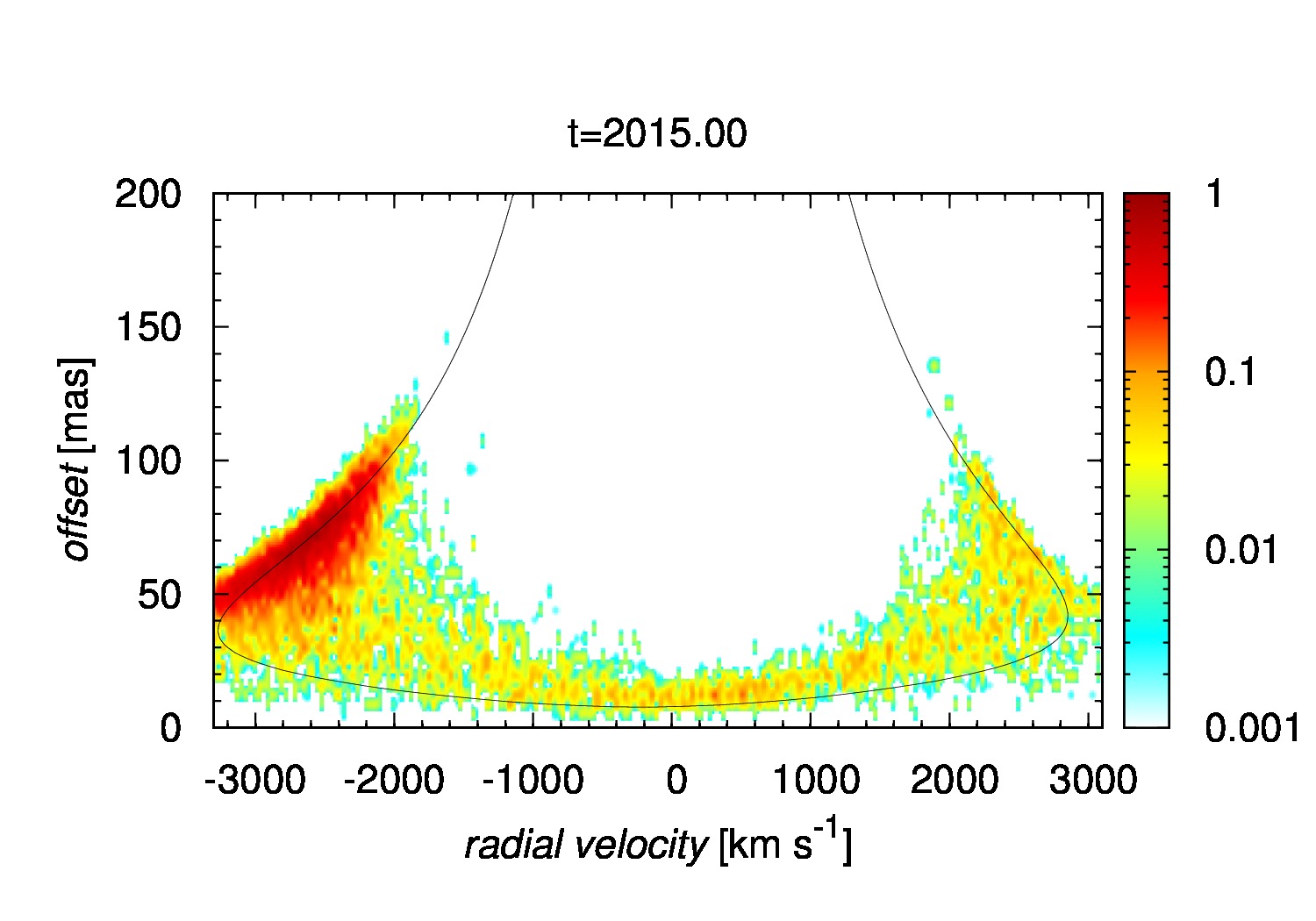} & \includegraphics[width=0.5\textwidth]{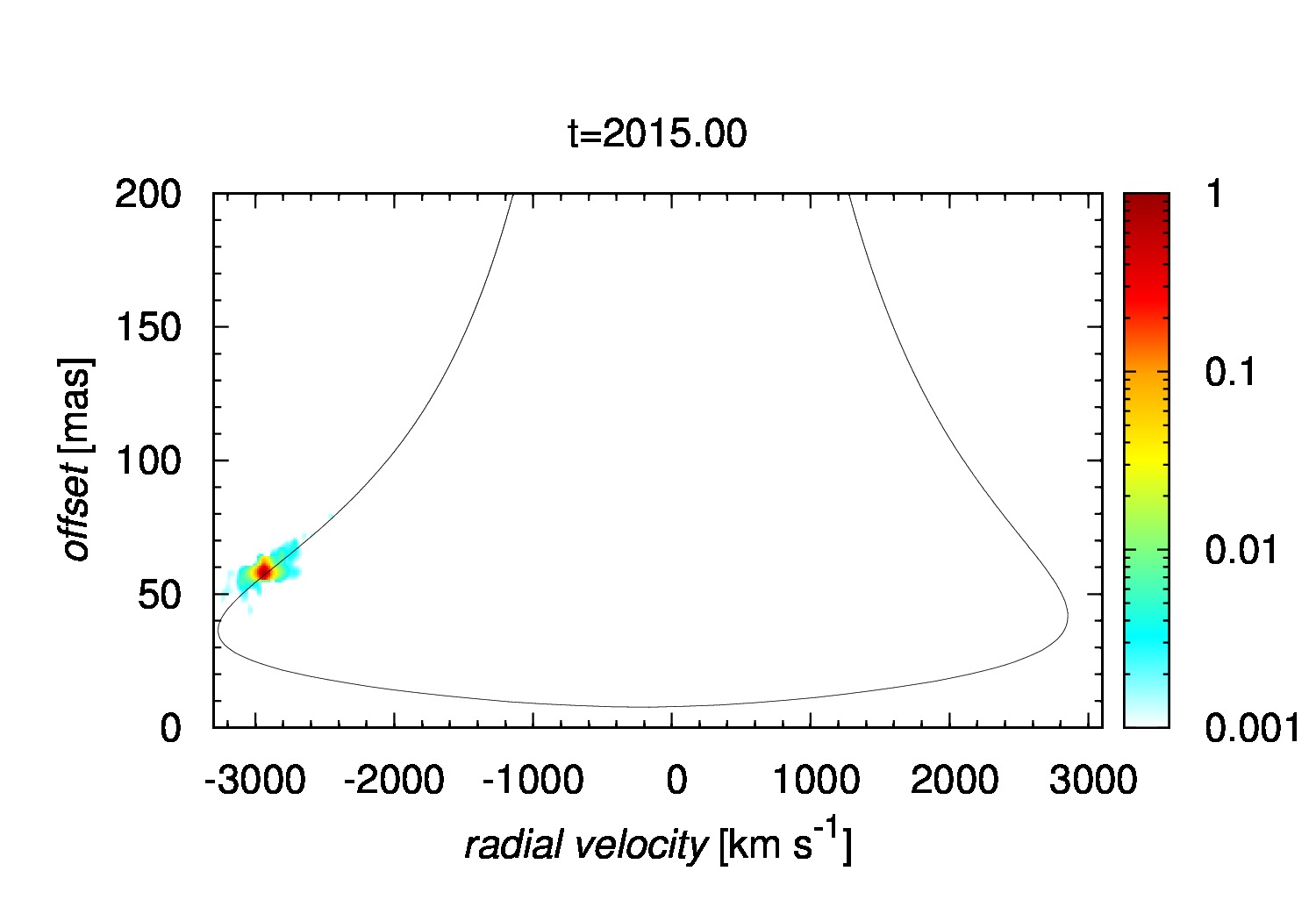}      
  \end{tabular}
  \caption{Position--velocity plots of a normalized density distribution (optically thin emission) for two scenarios of the DSO: \textbf{Left:} Core-less cloud. \textbf{Right:} A star of 1 $M_{\odot}$ surrounded by the circumstellar envelope. NIR observations \citep{val15} are consistent with the stellar scenario.}
  \label{img_simulation}
\end{figure}

Hence the observations are more consistent with the stellar scenario. To prove this, we perform numerical simulations of a star of 1 $M_{\odot}$ embedded in the circumstellar envelope whose density decreases as $r^{-2}$ and the initial velocity dispersion within the envelope is $50\,\rm{km\,s^{-1}}$. We see in Fig. \ref{img_simulation} (right panel) that the system star--envelope stays compact in the position--velocity plot (we use observable quantities for the plot -- offset from Sgr~A* in milliarcseconds and line-of-sight velocity in $\rm{km\,s^{-1}}$). The core-less cloud would get tidally stretched (see Fig. \ref{img_simulation}, left panel), leading to a simultaneous presence of redshifted and blueshifted emission close to the pericentre. The cloud should also be significantly prolonged near the pericentre by a factor of $\sim 3$--4, which was also not detected \citep{val15}. Thus, we consider an enshrouded star as our primary hypothesis for further observational tests.  

 \begin{figure}[tbh!]
      \minipage{0.45\textwidth}
      \includegraphics[width=\linewidth]{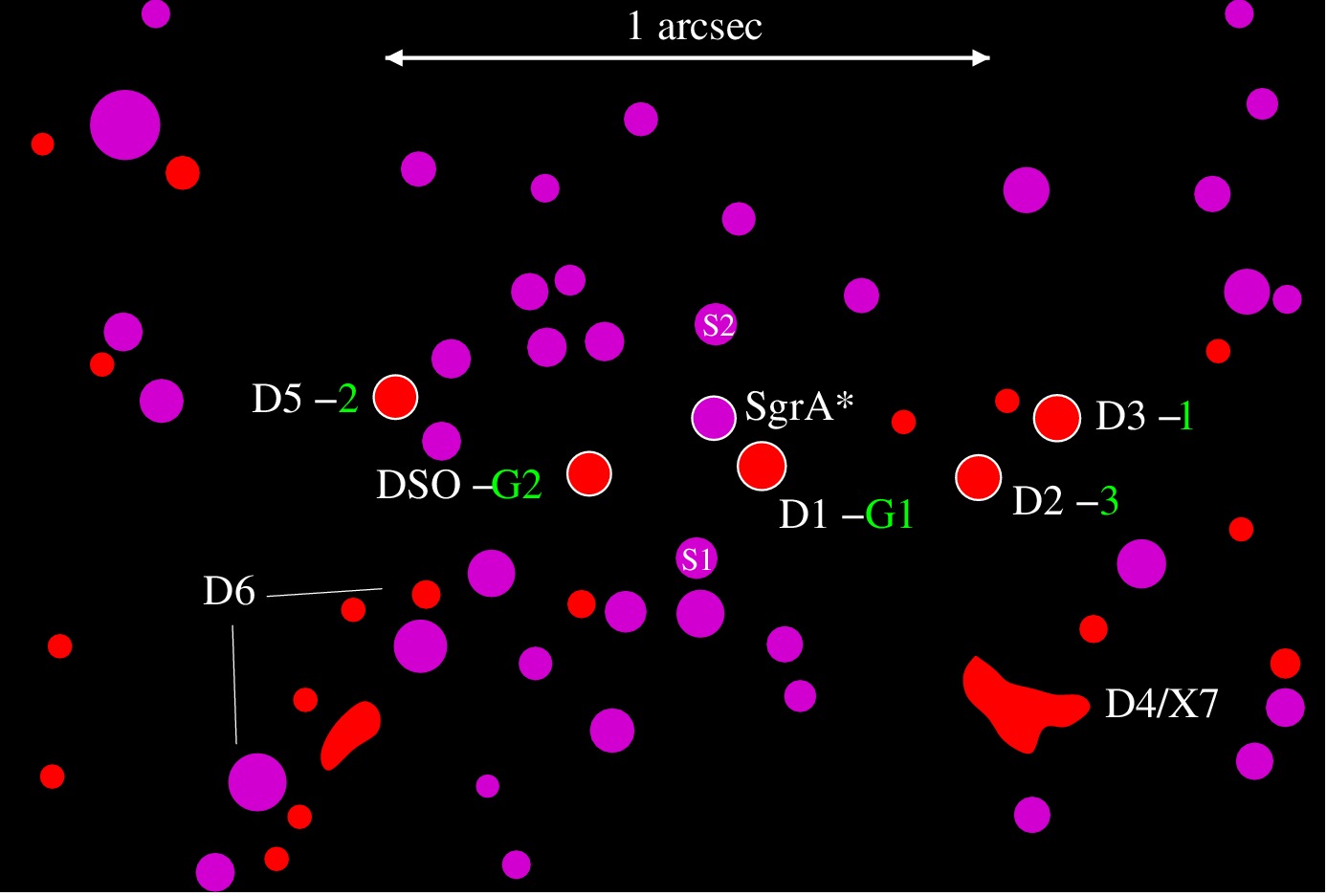}
      \endminipage\hfill
      \minipage{0.55\textwidth}
      \includegraphics[width=\linewidth]{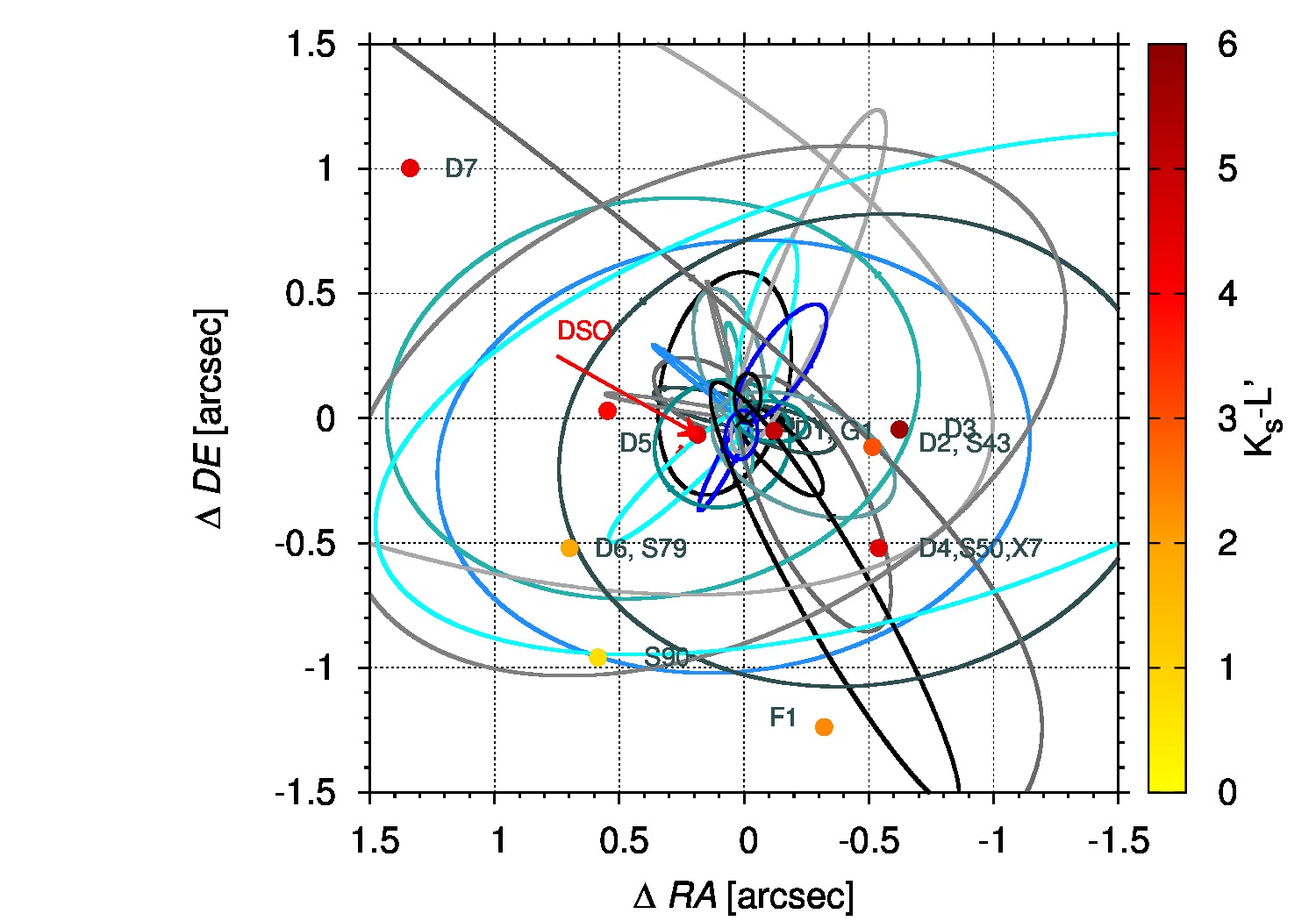}
      \endminipage\hfill
      \caption{\textit{Left:} Illustration of the position of infrared-excess sources near the Galactic centre as detected by \citet{eck13} (notation in white) and \citet{2014IAUS..303..264M} (notation in green). Figure adopted from \citet{2015arXiv150102164E}. \textit{Right:} Infrared excess $K_{s}-L'$ of detected sources as measured by \citet{eck13}. Stellar orbits of S-stars with orbital elements taken from \citet{gillessen09} are shown for illustration.}
  \label{fig_excess}
\end{figure}

The DSO belongs to a family of infrared-excess sources observed in the central $2.4'' \times 1.6''$ (see Fig. \ref{fig_excess} with positions in the S-cluster in the left panel and the corresponding difference in magnitudes in $K_{s}$ and $L'$ bands in the right panel). X7 source (also denoted as S50 or D4) is a known comet-shaped bow-shock source (see its analysis together with X3 source in \citet{muzic10}). Other sources can be pre-main-sequence stars with dusty envelopes/discs. They can also form bow shocks due to their supersonic motion through ambient medium that are not so well-resolved as X7 source.

\section{DSO as a young star}
\label{sec_star}

The width of the Br$\gamma$ emission line according to the analysis by \citet{val15} is $\sim 50 \pm 10\,$\AA\, before the pericentre and $15 \pm 10\,$\AA\,  after the pericentre, which corresponds to the line-of-sight Doppler contribution of $v_{{\rm LOS}}=\pm (294 \pm 60) {\rm km\,s}^{-1}$ and $v_{{\rm LOS}}=\pm (88 \pm 60) {\rm km\,s}^{-1}$, respectively. Regardless of the character of the DSO, line profiles can give us information about the kinematics of the emitting gas. Such large line widths of the order of $\sim 100\,{\rm km\,s}^{-1}$ can arise only from Doppler broadening due to the gas motion.

\begin{figure}[tbh]
  \centering
  \begin{tabular}{cc}
  \includegraphics[width=0.5\textwidth]{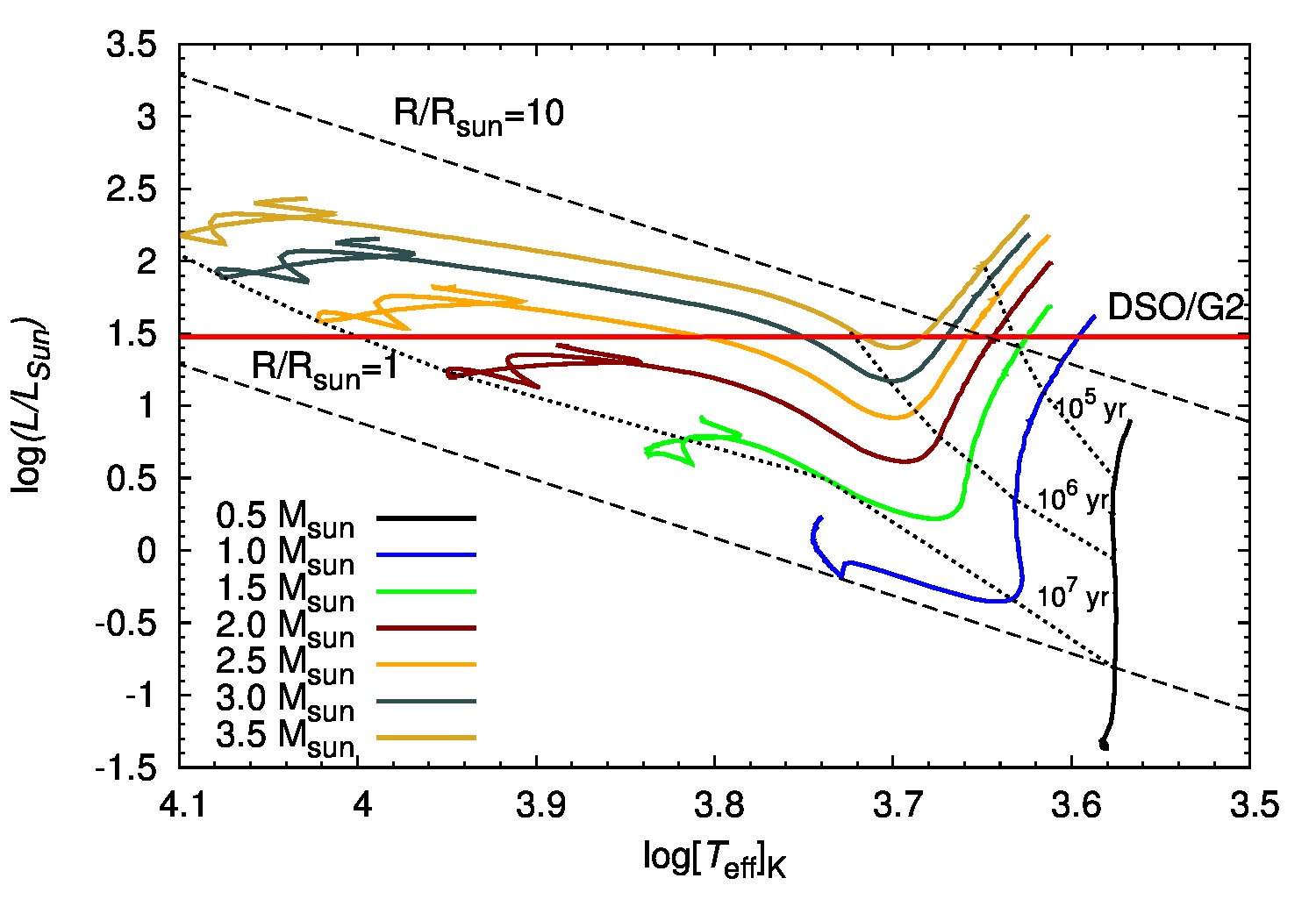} & \includegraphics[width=0.5\textwidth]{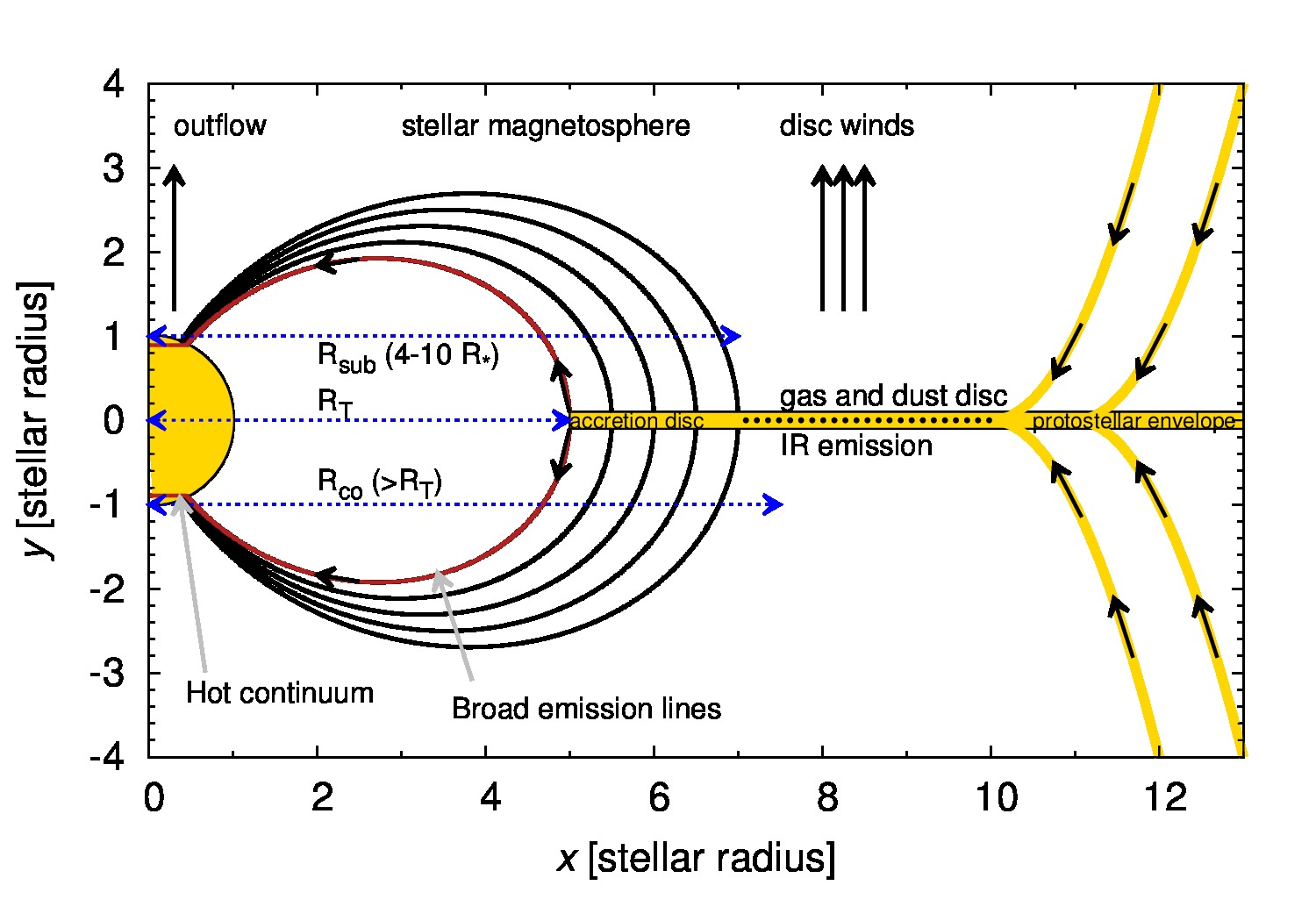} 
  \end{tabular}
   \caption{\textit{Left:} HR evolutionary diagram (luminosity--effective temperature) for several stars with different masses (see the legend). The upper limit on the luminosity of the DSO (30 $L_{\odot}$) is represented by a solid red line. Stellar radii as well as isochrones are represented by dashed lines.  Evolutionary tracks for pre-main-sequence stars were provided by \citet{Siess2000}. \textit{Right:} Schematic plot of magnetospheric model of pre-main-sequence stars that can explain broad emission line widths. Arrows show the length-scale of truncation radius, corotation radius, and sublimation radius.}
  \label{fig_evol}
\end{figure}

 In addition, \citet{val15} show that the bulk of the Br$\gamma$ emission comes from the compact region of less than $20\,\rm{mas}$ and only up to about 10$\%$ can be extended. Combined with the detection of only one prominent peak of Br$\gamma$ emission at each epoch, the character of the DSO is more consistent with a stellar scenario than a core-less cloud. Broad hydrogen recombination lines, which are observed in the spectrum of the DSO, are common among pre-main-sequence stars in the phase when they are accreting the matter from the circumstellar envelope that forms a disc closer to the star, see the right panel of Fig. \ref{fig_evol} for illustration. The Doppler contribution can cover the full range from $50\,\rm{km\,s^{-1}}$ up to $500\,\rm{km\,s^{-1}}$ \citep{bertout84}. 

The spectral decomposition of the DSO puts an upper limit on its bolometric luminosity, $L_{{\rm DSO}}\lesssim 30 \,L_{\odot}$. By analysing evolutionary tracks of pre-main-sequence-stars, see Fig. \ref{fig_evol} (left panel), the luminosity limit automatically constrains the mass as well as the radius of the putative star, $M_{{\rm DSO}}\lesssim 3\,M_{\odot}$, $R_{{\rm DSO}}\lesssim 10\,R_{\odot}$, and the most probable age lies in the range $10^5$--$10^6\,{\rm yr}$ when young stars are still in the phase of accretion.        

The line shape and luminosity of the observed Br$\gamma$ emission line may be reproduced within the \textit{magnetospheric accretion model} that is applied in \citet{val15} assuming axial symmetry (see also references therein, e.g. \citeauthor{bou07}, \citeyear{bou07} for a review). The basic assumption is that the putative star associated with the DSO is still accreting matter from the envelope/disc located within $\sim 0.1\,\rm{AU}$. Large infall velocities of the order of $100\,{\rm km\,s}^{-1}$ are reached by gas moving along magnetic streamlines from an inner portion of an accretion disc (see the right panel of Fig. \ref{fig_evol}). Pre-main-sequence stars exhibit a strong magnetic field of the order of 1$\,{\rm kG}$ that truncates the disc approximately at distance,

\begin{equation}
\frac{R_{\rm{T}}}{R_{\star}} \approx 6.5 B_{3}^{4/7} R_{2}^{5/7} \dot{M}_{-8}^{-2/7} M_{1}^{-1/7}\,,
\label{eq_truncation_radius}
\end{equation}   
where the strength of the dipole magnetic field at the equator of the star $B_{3}$ is in $\rm{kG}$, the stellar radius $R_{2}$ is in units of $2R_{\odot}$, the accretion rate $\dot{M}_{-8}$ is expressed in $10^{-8}M_{\odot}\rm{yr}^{-1}$, and the stellar mass $M_{1}$ is in units of $1 M_{\odot}$. The relation expressed by eq. \eqref{eq_truncation_radius} may serve as an upper limit for the disc truncation radius, since the ram gas presssure is higher for the disc geometry in comparison with the spherical geometry that was assumed in the derivation.    

Stable accretion proceeds only when the disc truncation radius is smaller that the corotation radius, which is defined as the radius where the Keplerian angular velocity is equal to the rotational angular velocity of the star and may be expressed in the following way,
\begin{equation}
R_{\rm{co}}\approx 4.2\, M_1^{1/3}P_{1}^{2/3}\,R_{\odot}, 
\end{equation}
where $M_{1}$ is stellar mass in units of $1\,M_{\odot}$ and $P_{1}$ is the stellar rotation period in units of $1$ day. 

The stellar radiation is reprocessed by irradiated discs which reemit in the near-, mid-, and far-infrared parts of spectrum due to presence of the dust beyond the dust sublimation radius (typically 4--10 stellar radii, see \citeauthor{val15}, \citeyear{val15} for details and Fig. \ref{fig_evol}). The gas from the inner portion of the disc is channelled along magnetic field lines and is shocked upon reaching the stellar surface, which gives rise to hot continuum (UV-, optical, and NIR-excess emission).

\paragraph*{Line and luminosity.} The Br$\gamma$ line luminosity is approximately of the order of $10^{-3}\,L_{\odot}$ and stays constant within uncertainties \citep[][see also Peissker et al., in preparation]{val15}. In the framework of the model of a pre-main-sequence star the line luminosity $L({\rm Br}\gamma)$ correlates with the accretion luminosity $L_{\rm acc}$ \citep{2014A&A...561A...2A},
 
 \begin{equation}
 \log{(L_{\rm{acc}}/L_{\odot})}=1.16 (0.07)\log{[L(\rm{Br} \gamma) / L_{\odot}]} + 3.60 (0.38)\,.
 \label{eq_correlation_luminosities}
 \end{equation}
For the DSO and its monitored Br$\gamma$ emission-line luminosity of $L(Br\gamma)= f_{\rm{acc}} \times 10^{-3}\, L_{\odot}$, where $f_{\rm{acc}}$ is a factor of the order of unity, we get the following values for the accretion luminosity, $(1.3,2.9,4.7,6.6)\,L_{\odot}$ for $f_{\rm{acc}}=\{1,2,3,4\}$. Hence, a fraction of the continuum emission of the DSO may be due to the hot continuum caused by the accretion. Given the accretion luminosity one can estimate the mass-accretion rate assuming the disc truncation radius of few stellar radii, e.g. $R_{\rm T}=5\,R_{\star}$,
    \begin{align}
  \dot{M}_{\rm{acc}} & \cong \frac{L_{\rm{acc}}R_{\star}}{GM_{\star}} \left(1-\frac{R_{\star}}{R_{\rm{T}}}\right)^{-1}\\
                     & \approx 4.1 \times 10^{-8} \left(\frac{L_{\rm{acc}}}{L_{\odot}}\right)\left(\frac{R_{\rm{star}}}{R_{\odot}}\right)\left(\frac{M_{\rm{star}}}{M_{\odot}}\right)^{-1}\,M_{\odot}\rm{yr^{-1}},
   \label{eq_mass_accretion_rate}
   \end{align}
which for the inferred mass and radius constraints,  $M_{{\rm DSO}}\lesssim 3\,M_{\odot}$ and $R_{{\rm DSO}}\lesssim 10\,R_{\odot}$, and the range of accretion luminosity leads to an estimate of $\dot{M}_{\rm acc}\lesssim 10^{-7}\,M_{\odot}{\rm yr}^{-1}$. Accretion is typically accompanied by outflows in young stellar systems, with the typical range of mass-loss rates $\dot{M}_{\rm w}/\dot{M}_{\rm acc} \approx 0.01-0.1$ \citep{edwards06}. 

 \begin{figure}[tbh]
   \begin{tabular}{cc}
      \includegraphics[width=0.5\textwidth]{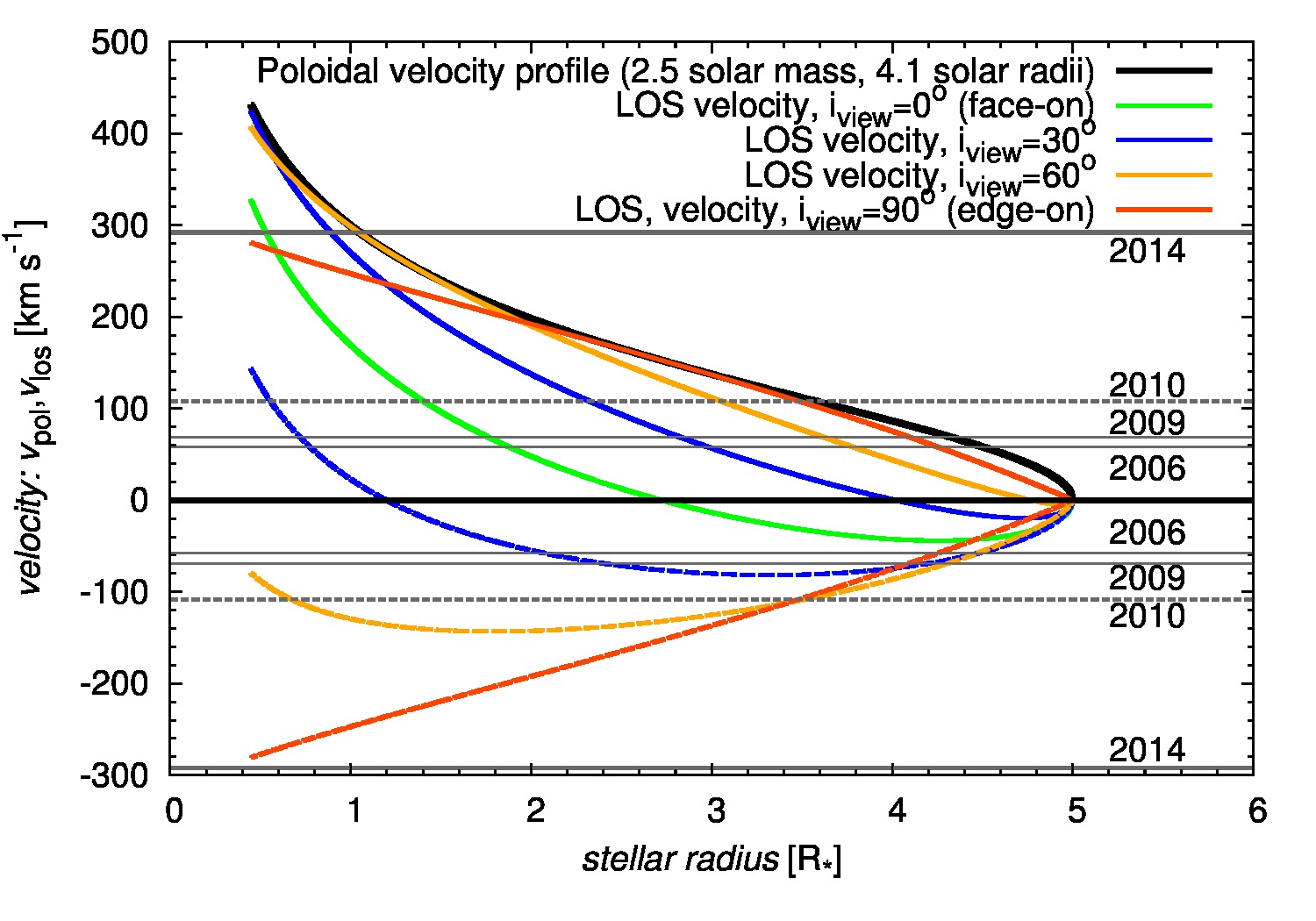} & \includegraphics[width=0.5\textwidth]{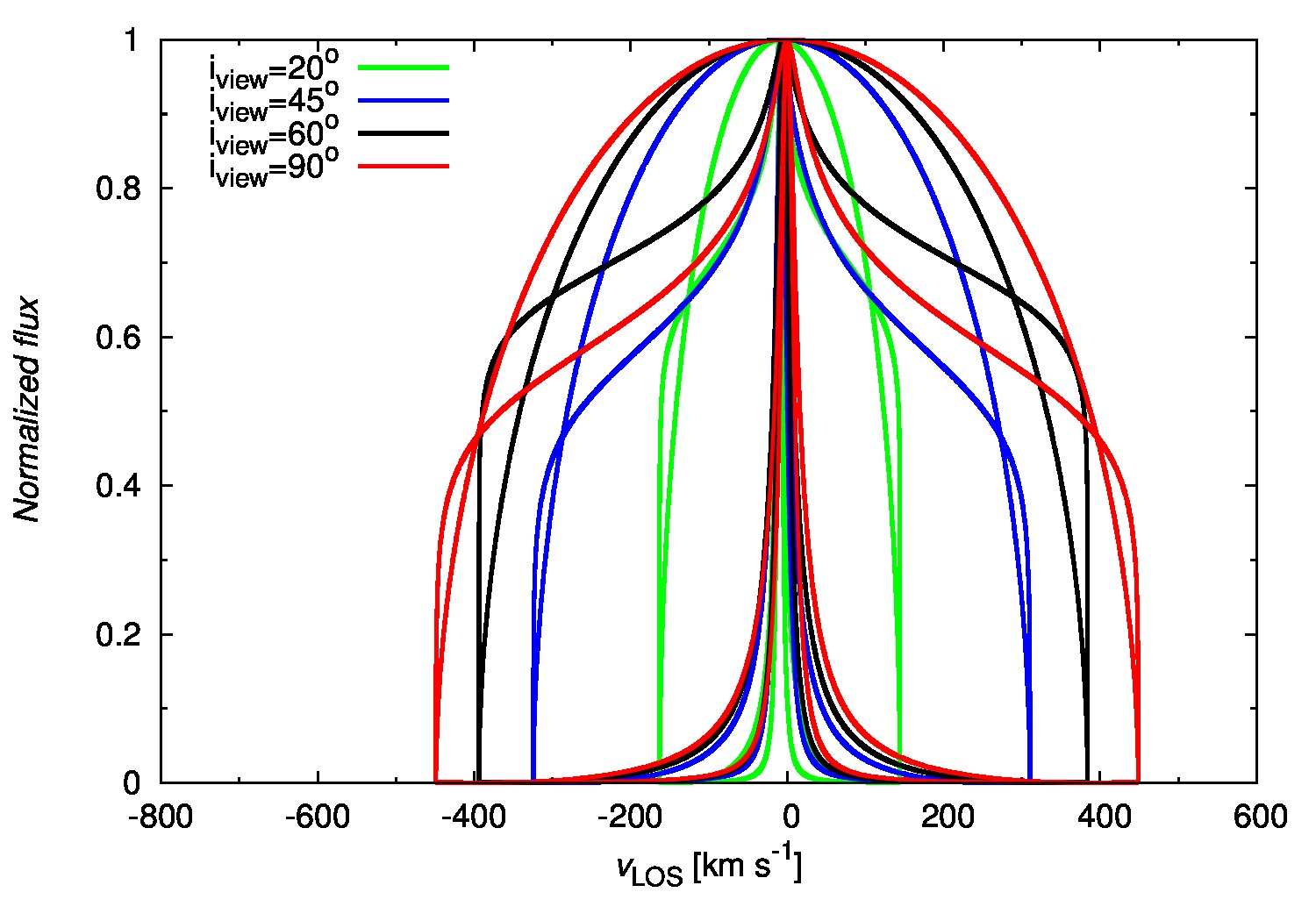}
   \end{tabular}
   \caption{\textit{Left:} The profile of maximum line-of-sight velocities of gas that is channelled along magnetic field lines.
   The distance is expressed in stellar radii. Different set of lines correspond to a different viewing angle. 
   \textit{Right:} Set of theoretical line profiles computed in axial symmetry. The line width increases for larger viewing angles, the maximum width is
   at $i_{{\rm view}}=90^{\circ}$. In our convention, the viewing angle of $90^{\circ}$ corresponds to the situation when we see the disc edge-on;
   the face-on view corresponds to zero degrees.
}
   \label{fig_linprof}
 \end{figure}

\paragraph*{Line width, shape, and variability.} A large line width of Br$\gamma$ line can be explained by magnetospheric accretion 
model that has been successful in explaining large widths of emission lines in other well-resolved young stellar objects \citep[see][for a review]{hartmann2000accretion}.
The basic concept is that the gas is channelled along magnetic field lines out of an inner part of the accretion disc onto the star. 
The estimate of poloidal velocities of accreting gas is as follows,

\begin{equation}
  v_{\rm pol} \approx \left[\frac{2GM_{\star}}{R_{\star}}\left(1-\frac{R_{\star}}{R_{\rm T}}\right)\right]^{1/2}\,.
  \label{eq_poloidal}
\end{equation}
The line-of-sight velocity that contributes to the Doppler broadening of emission lines depends naturally on the viewing angle of the star--disc system. 

 Finally, we compute the maximum line-of-sight velocities for the star having the mass of $2.5\,M_{\odot}$ that
 has a radius of $4.1\,R_{\odot}$ at the age of 1 Myr according to \citet{Siess2000}. The distance profile for
 different viewing angles is in the left-panel of Fig. \ref{fig_linprof}. The possible set of theoretical line profiles computed in axial
 symmetry according to \citet{1995ApJ...454..382K} is in the right panel of Fig. \ref{fig_linprof}. The observed line widths of Br$\gamma$ line \citep{val15} corresponding to several $100\,{\rm km\, s}^{-1}$ are
 reproduced well by a magnetospheric accretion model. There is also a considerable degree of variability corresponding to a different viewing angle. The line width increases for a larger viewing angle, 
 reaching the maximum width at $90^{\circ}$ (edge-on view).
 
  Hence, the observed variability of the width of Br$\gamma$ emission line of the DSO may be partially explained by the change of a viewing angle of the star-disc system as the source moves on a highly elliptical orbit around the SMBH. Other possible contributions include the perturbation of the disc/envelope by the tidal field of the SMBH, outflows (stellar wind, jet, disc winds), and bow-shock emission \citep{2016MNRAS.455.1257Z}.
  
  There are also sources of variability that are not intrinsic to the stellar system. These include the possible confusion due to the highly crowded region of the S-cluster \citep{val15}, which can also lead to the appearance of blend sources \citep{2012A&A...545A..70S}.


\section{Conclusions}
\label{conclusions}

The compact nature of the emission of the DSO (both continuum and line emission) supports the stellar character of this source. The observed near-infrared excess as well as the presence of emission lines imply the occurrence of the circumstellar material containing both gas and dust. The large line width of the monitored Br$\gamma$ emission can be explained by the Doppler broadening due to the gas that is channelled along magnetic field lines out of the inner part of the disc onto the star. The spectral decomposition combined with the model evolutionary tracks of young stars constrain the mass as well as the radius of the putative star ( $M_{{\rm DSO}}\lesssim 3\,M_{\odot}$ and $R_{{\rm DSO}}\lesssim 10\,R_{\odot}$). The confirmation of the presence of such a young stellar object (age $\sim 0.1$--$1\,{\rm Myr}$) so close to the supermassive black hole would support the theory of continuing star-formation in the Galactic centre.

\vspace*{0.5cm}

\acknowledgments 
Authors appreciate the hospitality of the Charles University in Prague, where this contribution was presented during the Week of Doctoral Students 2015. The  stay of M. Z., F. P., and G. D. K. in Prague was supported by the collaboration programme between the University of Cologne and the Charles University in Prague and the Czech Academy of Sciences--DAAD exchange programme between the University of Cologne and the Astronomical Institute of the Academy of Sciences in Prague.

\bibliographystyle{mymn2e.bst} 
\bibliography{zajacek} 




\end{article}

\end{document}